\documentclass[11pt]{article}
\usepackage{amsmath}
\usepackage{amssymb}
\usepackage{amscd}

\allowdisplaybreaks

\begin{document}

\begin{titlepage}
\thispagestyle{empty}
\begin{flushleft}
USCHEP/02-ib5
\hfill hep-th/0211131 \\
UT-02-58\hfill November, 2002 \\
\end{flushleft}

\vskip 1.5 cm
\bigskip

\begin{center}
\noindent{\Large \textbf{String Amplitudes from Moyal String Field Theory}}\\
\noindent{
 }\\
\renewcommand{\thefootnote}{\fnsymbol{footnote}}

\vskip 2cm
{\large
I. Bars$^{a,}$\footnote{e-mail address: bars@usc.edu},
I. Kishimoto$^{b,}$\footnote{e-mail address:
 ikishimo@hep-th.phys.s.u-tokyo.ac.jp} and
Y. Matsuo$^{b,}$\footnote{e-mail address:
 matsuo@phys.s.u-tokyo.ac.jp}, } \\
{\it
\noindent{ \bigskip }\\
$^{a)}$ Department of Physics and Astronomy,\\
University of Southern California, Los Angeles, CA 90089-0484, USA \\
\noindent{\smallskip  }\\
$^{b)}$ Department of Physics, Faculty of Science, University of Tokyo \\
Hongo 7-3-1, Bunkyo-ku, Tokyo 113-0033, Japan\\
\noindent{ \smallskip }\\
}
\bigskip
\end{center}
\begin{abstract}
We illustrate a basic framework for analytic computations of
Feynman graphs using the Moyal star formulation of string field
theory. We present efficient methods of computation based on (a)
the monoid algebra in noncommutative space and (b) the
conventional Feynman rules in Fourier space. The methods apply
equally well to perturbative string states or nonperturbative
string states involving D-branes. The ghost sector is formulated
using Moyal products with fermionic (b,c) ghosts. We also provide
a short account on how the purely cubic theory and/or VSFT
proposals may receive some clarification of their midpoint
structures in our regularized framework.
\end{abstract}
\vfill
\end{titlepage}\vfill\setcounter{footnote}{0} \renewcommand{\thefootnote}{%
\arabic{footnote}} \newpage


\section{Introduction}

During the past two years there has been a remarkable conceptual and
technical progress in string field theory (SFT) which was stimulated by its
application \cite{VSFT} to tachyon condensation, and the prospect of further
applications to more general physics of D-branes. The numerical computation
of the D-brane tension, for example, has reached a rather accurate estimate
\cite{Numerics}.

The role of SFT \cite{Witten} as a method to analyze non-perturbative string
phenomena has by now become rather evident. Consequently, efficient
computational tools to achieve analytic understanding of non-perturbative
string physics are now needed. Toward this goal, a new computational
technique has been developing over the past two years, starting with the
discovery \cite{B} of a direct connection between Witten's star product and
the usual Moyal star product that is well known in noncommutative geometry.
The new Moyal star $\star $ is applied on string fields $A\left( \bar{x}%
,x_{e},p_{e}\right) $ in the phase space of \textit{even} string modes,
independently for each $e.$ The product is local in the string midpoint $%
\bar{x}.$ Some basic numerical infinite matrices $T_{eo},R_{oe},w_{e},v_{o},$
\begin{equation}
T_{eo}=\frac{4o\left( i\right) ^{o-e+1}}{\pi \left( e^{2}-o^{2}\right) }%
,\;R_{oe}=\frac{4e^2\left( i\right) ^{o-e+1}}{\pi o\left( e^{2}-o^{2}\right)
},\;w_{e}=\sqrt{2}\left( i\right) ^{-e+2},\;v_{o}=\frac{2\sqrt{2}}{\pi }%
\frac{\left( i\right) ^{o-1}}{o}  \label{TRwv}
\end{equation}%
labeled by even/odd integers ( $e=2,4,6,\cdots ,$ and $o=1,3,5,\cdots $)
were needed to disentangle the Witten star into independent Moyal stars for
each mode $e.$ These matrices enter in a fundamental way in all string
computations in the Moyal star formulation of string field theory (MSFT).

In subsequent work \cite{BM1, BM2} MSFT was developed into as a precise
definition of string field theory, by resolving all midpoint issues,
formulating a consistent cutoff method in the number of string modes $2N$,
and developing a monoid algebra as an efficient and basic computational tool.

Computations in MSFT are based only on the use of the Moyal star product.
The new star provides an alternative to the oscillator tool or the conformal
field theory tool as a method of computation. In particular, cumbersome
Neumann coefficients or conformal maps that appear in the other approaches
to SFT are not needed, since they follow correctly from the Moyal star \cite%
{BM2}\footnote{%
Subsequent proposals of Moyal star products equivalent to the one in \cite{B}
have appeared \cite{DLMZ}\cite{Moyal}\cite{moyalMoscow}. They all become
discrete and well defined with the same cutoff method, and remain related to
the $\star $ which we use here.}.

A cutoff is needed in all formulations of SFT to resolve associativity
anomalies \cite{BM1}. The cutoff consists of working with a finite number of
string modes $n=1,2,\cdots ,2N$ that have oscillator frequencies $\kappa
_{n} $, and introducing finite $N\times N$ matrices $%
T_{eo},R_{oe},w_{e},v_{o}$ that are uniquely determined as functions of a
diagonal matrix $\kappa =diag\left( \kappa _{e},\kappa _{o}\right) $ which
represents arbitrary frequencies. The $\kappa _{n}=\left( \kappa _{e},\kappa
_{o}\right) $ are any reasonable functions of $n=\left( e,o\right) $,
including the possible choice of the usual oscillator frequencies $\kappa
_{n}=n$, even at finite $N $. The finite matrices $T,R,w,v$ are introduced
through the following defining relations (a bar means transpose)
\begin{equation}
R=\left( \kappa _{o}\right) ^{-2}\bar{T}\left( \kappa _{e}\right) ^{2},\quad
R=\bar{T}+v\bar{w},\quad v=\bar{T}w,\quad w=\bar{R}v\,.  \label{define}
\end{equation}%
The same relations are satisfied by the infinite matrices in Eq.(\ref{TRwv})
that have the usual frequencies $\kappa _{n}=n$ and $N\rightarrow \infty .$
These equations were uniquely solved in terms of arbitrary $\kappa _{n}$,$N$
\cite{BM2}$;$ 
\begin{eqnarray}
&&T_{eo}=\frac{w_{e}v_{o}\kappa _{o}^{2}}{\kappa _{e}^{2}-\kappa _{o}^{2}}%
,\quad R_{oe}=\frac{w_{e}v_{o}\kappa _{e}^{2}}{\kappa
_{e}^{2}-\kappa
_{o}^{2}}, \label{TR_exp}\\
&&w_{e}={i^{2-e}}\frac{\prod_{o^{\prime }}\left\vert \kappa
_{e}^{2}/\kappa _{o^{\prime }}^{2}-1\right\vert
^{\frac{1}{2}}}{\prod_{e^{\prime }\neq e}\left\vert \kappa
_{e}^{2}/\kappa _{e^{\prime}}^{2}-1\right\vert ^{\frac{1}{2}}},\quad
 v_{o}={i^{o-1}}
\frac{\prod_{e^{\prime }}\left\vert 1-\kappa _{o}^{2}/\kappa
_{e^{\prime }}^{2}\right\vert ^{\frac{1}{2}}}{\prod_{o^{\prime
}\neq o}\left\vert 1-\kappa _{o}^{2}/\kappa _{o^{\prime
}}^{2}\right\vert ^{\frac{1}{2}}}.\label{wv_exp}
\end{eqnarray}
For $\kappa _{n}=n$ and $N=\infty ,$ these reduce to the expressions in Eq.(%
\ref{TRwv}). Although the finite matrices are given quite explicitly, most
computations are done by using simple matrix relations among them without
the need for their explicit form. The following matrix relations are derived
\cite{BM2} from Eq.(\ref{define}):
\begin{eqnarray}
&&TR=1_{e},\quad RT=1_{o},\quad \bar{R}R=1+w\bar{w},\quad \bar{T}T=1-v\bar{v}%
,  \notag \\
&&T\bar{T}=1-\frac{w\bar{w}}{1+\bar{w}w},\quad Tv=\frac{w}{1+\bar{w}w},\quad
\bar{v}v=\frac{\bar{w}w}{1+\bar{w}w},  \label{relations} \\
&&Rw=v(1+\bar{w}w),\quad R\bar{R}=1+v\bar{v}\left( 1+\bar{w}w\right) .
\notag
\end{eqnarray}%
It is important to emphasize that in our formalism computing with arbitrary
frequencies $\kappa _{n}$ and finite number of modes $2N,$ is as easy as
working directly in the limit\footnote{%
The infinite matrices in Eq.(\ref{TRwv}) have well defined products when
multiplied two at a time, e.g. $TR=1_{e},$ $T\bar{T}=1_{e},$ etc.. However
they give ambiguous results in multiple matrix products due to associativity
anomalies \cite{BM1} that arise from marginally convergent infinite sums.
For example $(RT)v=v,$ but $R\left( Tv\right) =0.$ The unregulated Neumann
coefficients suffer from the same anomaly \cite{hata}\cite{BM2}. The finite
matrices resolve all ambiguities. One can follow how the anomaly occurs by
noting from Eq.(\ref{TRwv}) that $\bar{w}w\rightarrow \infty $ as $%
N\rightarrow \infty .$ For example, the zero in $Tv=w\left( 1+\bar{w}%
w\right) ^{-1}\rightarrow 0$ gets multiplied by an infinity that comes from
the product $Rw=v\left( 1+\bar{w}w\right) \rightarrow \infty .$ A unique
answer is obtained for any association, $RTv=v$, by doing all computations
at finite $N$ and taking the limit only at the end. \label{anom}}.

For example, as a test of MSFT, Neumann coefficients for any number of
strings were computed in \cite{BM2} with arbitrary oscillator frequencies $%
\kappa _{n}$ and cutoff $N.$ The cutoff version of Neumann coefficients $%
N_{mn}^{rs}\left( t\right) ,N_{0n}^{rs}\left( t,w\right) ,N_{00}^{rs}\left(
t,w\right) ,$ were found to be simple analytic expressions that depend on a
single $N\times N$ matrix $t_{eo}=\kappa _{e}^{1/2}T_{eo}\kappa _{o}^{-1/2}$
and an $N$-vector $w_e$. These explicitly satisfy the Gross-Jevicki nonlinear
relations for any $\kappa _{n},N$ \cite{BM2}. It is then evident that $T$
and $w$ (which follow from Eq.(\ref{define}) as functions of $\kappa $) are
more fundamental than the Neumann coefficients. As a corollary of this
result, by diagonalizing the matrix $t$ \cite{BM2} one can easily understand
at once why there is a Neumann spectroscopy for the 3-point vertex \cite%
{spectroscopy} or more generally the $n$-point vertex \cite{BM2}.

Such explicit analytic results, especially at finite $N,$ are new, and not
obtained consistently in any other approach. At finite $N$ the MSFT results
could be used in numerical as well as analytic computations as a more
consistent method than level truncation.

In this paper, we give a brief report on explicit computations of string
Feynman diagrams in MSFT. Related work, but in the oscillator formalism, is
pursued in \cite{Taylor}. Our formalism, with the finite $N$ regularization,
has the advantage that it applies in a straightforward manner when the
external states are either perturbative string states or non-perturbative
D-brane type states. So we can perform computations with the same ease when
nonperturbative states are involved. Our regularization plays a role similar
to that of lattice regularization in defining nonperturbative QCD. Any
string amplitude is analytically defined in this finite scheme. Furthermore,
we emphasize that to recover correctly the usual string amplitudes in the
large $N$ limit, it is essential that associativity anomalies are resolved
in the algebraic manipulations of $T,R,w,v$ in these computations$^{\ref%
{anom}}$. In this paper, we only present the basic ideas and the important
steps of the computation. The details will appear in a series of related
publications \cite{PREP}.

The organization of this paper is as follows. In section 2, we define the
regularized action for Witten's string field theory. In this paper, we will
work in the Siegel gauge where explicit realization of the finite $N$
regularization is possible. In section 3, we present Feynman diagram
computations in coordinate representation in noncommutative space. This is
an effective framework closely related to the methods in \cite{BM2}. In
section 4, we define systematically Feynman rules in the Fourier basis. This
is useful to see the connection with the conventional Feynman rules in
quantum field theory in noncommutative space \cite{Noncomm}. We present a
few examples of scattering amplitudes computed in both frameworks. In
section 5, we consider a re-organization of Feynman rules in Fourier space
to give a direct relation with the computations in section 3. In section 6,
we briefly outline the definition of Moyal product for the (fermionic) ghost
system. In section 7, we consider a possible relation with vacuum string
field theory (VSFT).

\section{Regularized action}

The starting point of our study is Witten's action \cite{Witten} for the
open bosonic string, taken in the Siegel gauge, and rewritten in the Moyal
basis
\begin{equation}
S=\int d^{d}\bar{x}\,Tr\left( \frac{1}{2}A\star (L_{0}-1)A+\frac{1}{3}A\star
A\star A\right) \,\,.  \label{action}
\end{equation}%
The field $A\left( \bar{x},\xi \right) $ depends on the noncommutative
coordinates $\xi _{i}^{\mu }=\left( x_{2}^{\mu },x_{4}^{\mu }\cdots
,x_{2N}^{\mu },p_{2}^{\mu },p_{4}^{\mu }\cdots ,p_{2N}^{\mu }\right) $. The $%
\xi _{i}^{\mu }$ may include ghosts in either the bosonic or the fermionic
version. The bosonized ghost was discussed in \cite{BM2} as a 27'th bosonic
coordinate $\xi _{i}^{27}=\left( \phi _{e},\pi _{e}\right) $, while the
fermionic case will be discussed in a later section in this paper. In the
following sections, however, we concentrate basically on the matter sector
for the simplicity of argument. The Moyal product $\star $ and the trace $Tr$
are defined at fixed $\bar{x}$ as,\footnote{%
In the following, we denote $d^{d}\xi _{1}\cdots d^{d}\xi _{2N}$ as $(d\xi )$.%
}
\begin{equation}
(A\star B)(\bar{x},\xi )=A(\bar{x},\xi )~e^{\frac{1}{2}\eta _{\mu \nu }%
\overleftarrow{\partial _{\mu }^{i}}\sigma _{ij}\overrightarrow{\partial
_{\nu }^{j}}}B(\bar{x},\xi ),\;Tr\left( A\left( \bar{x}\right) \right) =\int
\frac{\left( d\xi \right) }{\left( \det 2\pi \sigma \right) ^{d/2}}A\left(
\bar{x},\xi \right) .  \label{Moyal}
\end{equation}%
The string field lives in the direct product of the Moyal planes, with $%
\left[ \xi _{i}^{\mu },\xi _{j}^{\nu }\right] _{\star }=\sigma _{ij}\,\eta
^{\mu \nu },$ where
\begin{equation}
\sigma _{ij}=i\theta \left(
\begin{array}{cc}
0 & 1_{e} \\
-1_{e} & 0%
\end{array}%
\right) .
\end{equation}%
The parameter $\theta $ absorbs units and could be mapped to 1 by a
rescaling of the units of $p_{e}.$

The kinetic term is given by the Virasoro operator $L_{0}$ which was
computed in Moyal space in \cite{BM2}. Here we rewrite it in the form of a
differential operator%
\begin{equation}
L_{0}=\frac{1}{2}\beta _{0}^{2}-\frac{d}{2}Tr\left( \tilde{\kappa}\right) -%
\frac{1}{4}\bar{D}_{\xi }\left( M_{0}^{-1}\tilde{\kappa}\right) D_{\xi }\,+%
\bar{\xi}\left( \tilde{\kappa}M_{0}\right) \xi ,\;\;  \label{Lo}
\end{equation}%
where $\beta _{0}=-il_{s}\frac{\partial }{\partial \bar{x}}$, $D_{\xi
}=\left( \left( \frac{\partial }{\partial x_{e}}-i\frac{\beta _{0}}{l_{s}}%
w_{e}\right) ,\;\frac{\partial }{\partial p_{e}}\right) ,$ and $l_{s}$ is
the string length. The $\left( 2N\right) \times \left( 2N\right) $ matrices
\begin{equation}
\;\tilde{\kappa}=\left(
\begin{array}{cc}
\kappa _{e} & 0 \\
0 & T\kappa _{o}R%
\end{array}%
\right) ,\;M_{0}=\left(
\begin{array}{cc}
\frac{\kappa _{e}}{2l_{s}^{2}} & 0 \\
0 & \frac{2l_{s}^{2}}{\theta ^{2}}T\kappa _{o}^{-1}\bar{T}%
\end{array}%
\right) ,  \label{Mo}
\end{equation}%
give the block diagonal forms $M_{0}^{-1}\tilde{\kappa}=diag\left(
2l_{s}^{2}1_{e},\frac{\theta ^{2}}{2l_{s}^{2}}\kappa _{e}^{2}\right) $ and $%
\tilde{\kappa}M_{0}=diag\left( \frac{\kappa _{e}^{2}}{2l_{s}^{2}},\frac{%
2l_{s}^{2}}{\theta ^{2}}\left( T\bar{T}\right) _{ee^{\prime }}\right) $
after using Eqs.(\ref{define},\ref{relations}). We note that $T\bar{T}$ in $%
\tilde{\kappa}M_{0}$ is almost diagonal, since $T\bar{T}=1-\frac{w\bar{w}}{1+%
\bar{w}w}$, and the second term becomes naively negligible in the large $N$
limit since $\bar{w}w\rightarrow \infty $. A major simplification would
occur if one could neglect this term. However, with this simplification one
cannot recover the string one-loop amplitude or other quantities correctly,
as we will see below in Eq.(\ref{oneloop}), because of the anomalies
discussed in footnote (\ref{anom}). The lesson is that one should not take
the large $N$ limit at the level of the Lagrangian\footnote{%
This term in $L_{0}$ was missed in \cite{DLMZ} in their attempt to compare
the discrete Moyal $\star _{e}$ of \cite{B} to the continuous Moyal $\star
_{\kappa }$ directly at $N=\infty ,$ and erroneously concluded that there
was a discrepancy. In fact, there is full agreement. \label{wrong}}. One
should do it only after performing all the algebraic manipulations that
define the string diagram. Consequently all of the following expressions are
at finite $N$ unless specified otherwise.

The string field that represents the perturbative vacuum is given by the
gaussian $A_{0}\sim \exp \left( -\bar{\xi}M_{0}\xi \right) $ (for any $%
\kappa _{n},N$). The on-shell tachyon state $\left( L_{0}-1\right) A_{t}=0$
is given by $A_{0}e^{ik\cdot x_{0}},$ which is
\begin{equation}
A_{t}\left( \bar{x},\xi \right) =\mathcal{N}_{0}e^{-\bar{\xi}M_{0}\xi -\bar{%
\xi}\lambda _{0}}e^{ik\cdot \bar{x}},\;\left( \lambda _{0}\right) _{i}^{\mu
}=-ik^{\mu }\left( w_{e},0\right) ,\;\mathcal{N}_{0}=\left( \det 4\sigma
M_{0}\right) ^{d/4},\;l_{s}^{2}k^{2}=2\,.  \label{tach}
\end{equation}%
The form of $\left( \lambda _{0}\right) _{i}^{\mu }$ for the tachyon follows
from $e^{ik\cdot x_{0}}$ after rewriting the center of mass coordinate $%
x_{0} $ in terms of the midpoint $\bar{x},$ i.e. $x_{0}=\bar{x}+w_{e}x_{e}.$
The norm $\mathcal{N}_{0}$ is fixed by requiring $Tr\left( A_{t}^{\ast
}\star A_{t}\right) =1.$

All perturbative string states with definite center of mass momentum $k^{\mu
}$ are represented by polynomials in $\xi $ multiplying the tachyon field.
All of them can be obtained from the following generating field by taking
derivatives with respect to a general $\lambda $
\begin{equation}
A\left( \bar{x},\xi \right) ={\cal N} e^{-\bar{\xi}M_{0}\xi -\bar{\xi}\lambda}~e^{ik\cdot \bar{x}},
\end{equation}%
and setting $\lambda \rightarrow \lambda _{0}=-ik^{\mu }\left(
w_{e},0\right) $ at the end.

Nonperturbative string fields that describe D-branes involve projectors in
VSFT conjecture \cite{VSFT}. A general class is \cite{BM2}
\begin{equation}
A_{D,\lambda }\left( \xi \right) =\mathcal{N}\exp \left( -\bar{\xi}D\xi -%
\bar{\xi}\lambda \right) ,\;\mathcal{N}=2^{dN}\exp (-\frac{1}{4}\bar{\lambda}%
\sigma D\sigma \lambda ),\;D=\left(
\begin{array}{cc}
a & ab \\
ba & ~\frac{1}{a\theta^2}+bab%
\end{array}%
\right)\,.
\end{equation}%
For any $\lambda ,$ and symmetric $a,b,$ these satisfy $A\star A=A,$ and $%
TrA=1.$ For a D-brane the components of $\lambda $ parallel to the brane
vanish, $\lambda ^{\parallel }=0,$ while those perpendicular to the brane
are nonzero as a function of the midpoint $\lambda ^{_{\bot }}\left( \bar{x}%
_{\bot }\right) \neq 0.$ Examples such as the sliver field, butterfly field,
etc., are special cases of these formulas with specific forms of the matrix $%
D$ \cite{BM2}.

It appears that for all computations of external states of interest we
should consider the field configurations that contain the general parameters
${\cal N},M_{ij},\lambda _{i}^{\mu },k^{\mu }$%
\begin{equation}
A_{\mathcal{N},M,\lambda ,k}=\mathcal{N}\exp \left( -\bar{\xi}M\xi -\bar{\xi}%
\lambda +ik\cdot \bar{x}\right) ,  \label{external}
\end{equation}%
where for perturbative states $\mathcal{N}$ is a constant, but for D-brane
states it may depend on $\bar{x}$. These fields form a closed algebra under
the star%
\begin{eqnarray}
&&\left( \mathcal{N}_{1}\exp \left( -\bar{\xi}M_1\xi -\bar{\xi}\lambda_1
+ik_{1}\cdot \bar{x}\right) \right) \star \left( \mathcal{N}_{2}\exp \left( -%
\bar{\xi}M_{2}\xi -\bar{\xi}\lambda _{2}+ik_{2}\cdot \bar{x}\right) \right)
\\
&=&\left( \mathcal{N}_{12}\exp \left( -\bar{\xi}M_{12}\xi -\bar{\xi}\lambda
_{12}+i\left( k_{1}+k_{2}\right) \cdot \bar{x}\right) \right)   \label{mono12}
\end{eqnarray}%
where the structure of $\mathcal{N}_{12},\left( M_{12}\right) _{ij},\left(
\lambda _{12}\right) _{i}^{\mu }$ is given as (define $m_{1}=M_{1}\sigma
,\;m_{2}=M_{2}\sigma ,\;m_{12}=M_{12}\sigma $)
\begin{eqnarray}
m_{12} &=&\left( m_{1}+m_{2}m_{1}\right) \left( 1+m_{2}m_{1}\right)
^{-1}+\left( m_{2}-m_{1}m_{2}\right) \left( 1+m_{1}m_{2}\right) ^{-1},
\label{m12} \\
\lambda _{12} &=&\left( 1-m_{1}\right) \left( 1+m_{2}m_{1}\right)
^{-1}\lambda _{2}+\left( 1+m_{2}\right) \left( 1+m_{1}m_{2}\right)
^{-1}\lambda _{1},  \label{lambda12} \\
\mathcal{N}_{12} &=&\frac{\mathcal{N}_{1}\mathcal{N}_{2}}{\det \left(
1+m_{2}m_{1}\right) ^{d/2}}e^{\frac{1}{4}\left( \left( \bar{\lambda}_{1}+%
\bar{\lambda}_{2}\right) \sigma \left( m_{1}+m_{2}\right) ^{-1}\left(
\lambda _{1}+\lambda _{2}\right) -\bar{\lambda}_{12}\sigma \left(
m_{12}\right) ^{-1}\lambda _{12}\right) }\,\,.  \label{n12}
\end{eqnarray}%
This algebra is a monoid, which means it is associative, closed, and
includes the identity element (number 1). It is short of being a group since
some elements (in particular projectors) do not have an inverse, although
the generic element does have an inverse. The trace of a monoid is given
through Eq.(\ref{Moyal}) (assuming a decaying exponential in $\xi $)%
\begin{equation}
Tr\left( A_{\mathcal{N},M,\lambda ,k}\right) =\frac{\mathcal{N}e^{ik\cdot
\bar{x}}e^{\frac{1}{4}\bar{\lambda}M^{-1}\lambda }}{\det \left( 2M\sigma
\right) ^{d/2}}.  \label{trace}
\end{equation}%
Building on the computations in \cite{BM2}, this monoid algebra will be used
as a basic computational tool in evaluating string field theory diagrams.

\section{Feynman graphs in $\protect\xi $ basis}

In this section we discuss Feynman graphs in the noncommutative $\xi $ basis
and in the next section we formulate them in the Fourier transformed basis.

In a Feynman diagram an external string state will be represented by a
monoid $A\left( \bar{x},\xi \right) $ that corresponds to a perturbative
string state or nonperturbative D-brane state as discussed in the previous
section. This corresponds to a boundary condition in the language of a
worldsheet representation of the Feynman diagram. The propagator is given as
an integral using a Schwinger parameter $\left( L_{0}-1\right)
^{-1}=\int_{0}^{\infty }d\tau e^{\tau }\exp \left( -\tau L_{0}\right) .$
This corresponds to the free propagation of the string as represented by the
worldsheet between the boundaries.

To evaluate Feynman diagrams we will need the $\tau $-evolved monoid element
\begin{equation}
e^{-\tau L_{0}}\left( \mathcal{N}e^{-\bar{\xi}M\xi -\bar{\xi}\lambda
}e^{ip\cdot \bar{x}}\right) =\mathcal{N}\!\left( \tau \right) ~e^{-\bar{\xi}%
M\left( \tau \right) \xi -\bar{\xi}\lambda \left( \tau \right) }e^{ip\cdot
\bar{x}}.  \label{evolution}
\end{equation}%
Both sides must be annihilated by the Schr\"{o}dinger operator $\left(
\partial _{\tau }+L_{0}\right) .$ The result of the computation is given by%
\begin{align}
M\!\left( \tau \right) & =\left[ \sinh\tau \tilde{\kappa} +\left( \sinh\tau
\tilde{\kappa} +M_{0}M^{-1}\cosh \tau \tilde{\kappa}\right) ^{-1}\right]
\left( \cosh \tau \tilde{\kappa}\right) ^{-1}M_{0} ,  \label{tM} \\
\lambda\!\left( \tau \right) & =\left[ \left( \cosh \tau \tilde{\kappa}%
+MM_{0}^{-1}\sinh \tau \tilde{\kappa}\right) ^{-1}\left( \lambda +iwp\right) %
\right] -iwp ,  \label{tlam} \\
\mathcal{N}\!\left( \tau \right) & =\frac{\mathcal{N}~e^{-\frac{1}{2}%
l_{s}^{2}p^{2}\tau }~\exp \left[ \frac{1}{4}\left( \bar{\lambda}+ip\bar{w}%
\right) \left( M+\coth \tau \tilde{\kappa}~M_{0}\right) ^{-1}\left( \lambda
+iwp\right) \right] }{\det \left( \frac{1}{2}\left( 1+MM_{0}^{-1}\right)
+\frac{1}{2}\left( 1-MM_{0}^{-1}\right)e^{-2\tau \tilde{\kappa}} \right)
^{d/2}} .  \label{tN}
\end{align}%
For the tachyon in Eq.(\ref{tach}) this simplifies greatly $M_{0}\left( \tau
\right) =M_{0},$ $\lambda _{0}\left( \tau \right) =\lambda _{0},$ $\mathcal{N%
}_{0}\left( \tau \right) =\mathcal{N}_{0}e^{-\tau }.$ Note that even in the
general case the evolved monoid is also a monoid that can be star multiplied
easily with other monoids.

The diagrams below will be given as a function of the Schwinger parameters $%
\tau _{i}$. The function should be integrated using the measure $%
\int_{0}^{\infty }d\tau _{i}e^{\tau _{i}}$ for each propagator. We now give
some examples of tree diagrams

\begin{itemize}
\item The diagram $1-2$ for two external states $A_{1},A_{2}$ joined by a
propagator is given by
\begin{eqnarray}
&&\int d^{d}\bar{x}\,Tr\left( A_{1}\star e^{-\tau L_{0}}A_{2}\right)   \notag
\\
&=&\frac{\mathcal{N}_{1}\mathcal{N}_{2}\!\left( \tau \right) \exp \left(
\frac{1}{4}\left( \bar{\lambda}_{1}+\bar{\lambda}_{2}\!\left( \tau \right)
\right) \left( M_{1}+M_{2}(\tau )\right) ^{-1}\left( \lambda _{1}+\lambda
_{2}\!\left( \tau \right) \right) \right) }{\left( \det \left( 2\left(
M_{1}+M_{2}\left( \tau \right) \right) \sigma \right) \right) ^{d/2}}\left(
2\pi \right) ^{d}\delta ^{d}\left( p\right)   \label{prop12}
\end{eqnarray}%
where $p^{\mu }=k_{1}^{\mu }+k_{2}^{\mu }$. To evaluate it we used Eqs.(\ref%
{evolution}-\ref{tN}), and the trace in Eq.(\ref{trace}). For tachyon
external states of Eq.(\ref{tach}) this expression collapses to just $%
e^{-\tau }\left( 2\pi \right) ^{d}\delta ^{d}\left( p\right) $, which is the
expected result. For more general states our formula provides an explicit
analytic result.

\item The 4 point function is computed from the diagram for $_{1}^{2}{>}-{<}%
_{4}^{3}$ and its various permutations of $\left( 1,2,3,4\right) .$ The MSFT
expression for this diagram is
\begin{equation}
_{12}A_{34}=\int d^d\bar{x}\,Tr\left( e^{-\tau L_{0}}\left( A_{1}\star
A_{2}\right) \star A_{3}\star A_{4}\right).  \label{four}
\end{equation}%
The two external lines $\left( 1,2\right) $ are joined to the resulting
state by the product $A_{12}=A_{1}\star A_{2},$ which is a monoid as given
in Eq.(\ref{mono12}). This monoid is propagated to $A_{12}\left( \tau
\right) =e^{-\tau L_{0}}\left( A_{1}\star A_{2}\right) $ by using Eqs.(\ref%
{evolution}-\ref{tN}), and then traced with the monoid
$A_{34}=A_{3}\star A_{4}.$
 Then the computation of the four point function is completed by
using the formula in Eq.(\ref{prop12}). That is, replace the
monoid $A_1$ by the monoid $A_{34}$, and similarly $A_2$ by
$A_{12}$, and apply Eqs.({\ref{tM}}-{\ref{tN}}) for the monoid
$A_{12}$. The remainder of the computation is straightforward
algebra and those details will be published in a future
paper.\cite{PREP}  We emphasize that the external states can be
nonperturbative. For the case of perturbative tachyon scattering,
for {\it off shell tachyons}, the result is
\begin{eqnarray}
{}_{12}A_{34} &=&\frac{\det \left( 2m_{0}\right) ^{d/2}}{\det \left(
1+m_{0}^{2}\right) ^{d}}(2\pi )^{d}\delta
^{d}(p_{1}+p_{2}+p_{3}+p_{4})
\notag \\
&&\times \frac{e^{- \frac{1}{2}l_{s}^{2}(p_{1}+p_{2})^{2}(\tau
+\alpha (\tau ))}~e^{l_{s}^{2}(p_{1}+p_{3})^{2}\beta (\tau
)}~e^{\frac{1}{2}l_{s}^{2}\sum_{i=1}^4 p_{i}^{2}\gamma (\tau
)}}{(\det (2G_{e}(\tau )e^{\tau \kappa _{e}}))^{-d/2}(\det
(2G_{o}(\tau )e^{\tau \kappa _{o}}))^{-d/2}} \label{4-amp_rev}
\end{eqnarray}
where
\begin{eqnarray}
\alpha \left( \tau \right)  &=&\bar{z}\left[ \bar{t}G_{e}\left(
E_{e}-1\right) t+G_{o}\left( E_{o}-1\right) \right] z, \\
\beta \left( \tau \right)  &=&\bar{z}G_{o}z,\;\;\gamma \left( \tau
\right) = \bar{z}G_{o}\left( E_{o}-1\right) z-\bar{z}(1+\bar{t}t)z
\end{eqnarray}
are given in terms of the following definitions
\begin{eqnarray}
&&z =\left( 1+\bar{t}t\right) ^{-1}\bar{t}\kappa _{e}^{-1/2}w,\;\;t=\kappa
_{e}^{1/2}T\kappa _{o}^{-1/2},\;\;m_{0}=M_{0}\sigma\,,  \\
&&E_{e}\left(\tau\right)=\cosh \left( \kappa _{e}\tau
\right) +\frac{2}{1+t\bar{t}}\sinh (\kappa _{e}\tau ),\quad
E_{o}\left( \tau \right)=\cosh \left( \kappa _{o}\tau
\right) +\frac{2}{1+\bar{t}t}\sinh (\kappa _{o}\tau )\,, \\
&&S_{e,o}\left( \tau \right)  =\sinh (\kappa _{e,o}\tau ),\quad
G_{e,o}\left( \tau \right) =2S_{e,o}\left( E_{e,o}^{2}-1\right) ^{-1}\,.
\end{eqnarray}
Before integrating with the measure
  $\int_0^{\infty} d\tau e^{\tau}$
we also need to multiply this expression by the ghost
contribution, which will appear in our future paper. This should
reproduce the Veneziano formula when all tachyons are put on shell
$l_s^2 p^2_i =2$, and we take the large $N$ limit with
$\kappa_n=n$.

\item Similarly, the diagram $_{1}^{2}{>}-^{^{3}|}-{<}_{5}^{4}$ involves
\begin{equation}
\int d^d\bar{x}\,Tr\left( A_{12}\left( \tau _{1}\right) \star A_{3}\star
A_{45}\left( \tau _{2}\right) \right) =\int d^d\bar{x}\,Tr\left( e^{-\tau
_{1}L_{0}}\left( A_{1}\star A_{2}\right) \star A_{3}\star e^{-\tau
_{2}L_{0}}\left( A_{4}\star A_{5}\right) \right) .
\end{equation}

\item The diagram $_{1}^{2}{>}-^{^{3}|}-^{^{4}|}-{<}_{6}^{5}$ involves
\begin{equation}
\int d^d\bar{x}\,Tr\left( A_{12}\left( \tau _{1}\right) \star A_{3}\star
e^{-\tau _{2}L_{0}}\left( A_{4}\star A_{56}\left( \tau _{3}\right) \right)
\right) .
\end{equation}
\end{itemize}

Next we consider loop diagrams. We start from an expression given above for
a tree diagram and then identify any two external lines to make a closed
loop. Suppose the external legs that are identified were represented by the
fields $A_{i},A_{j}$ in the tree diagram. In the loop these fields are
replaced by
\begin{equation}
A_{i}\rightarrow e^{-\tau _{i}L_{0}}\left( e^{i\bar{\xi}\eta }e^{ip\cdot
\bar{x}}\right) ,\;\;A_{j}\rightarrow e^{-i\bar{\xi}\eta }e^{-ip\cdot \bar{x}%
}
\end{equation}%
and the integral over $\eta $ is performed (the Fourier basis is a complete
set of states to sum over in the propagation). Here $\tau _{i}$ is the
modulus of the new propagator and $p^{\mu }$ becomes the momentum flowing in
this propagator by momentum conservation. Some examples of loops follow

\begin{itemize}
\item The one loop diagram $\bigcirc $ with no external legs is obtained
from the 2-point vertex $Tr\left( A_{1}\star A_{2}\right) $ by identifying
legs 1,2. This leads to the integral
\begin{equation}
\int d^d\bar{x}\int\! \frac{d^dp}{\left( 2\pi \right) ^{d}}\frac{(d\eta) }{%
\left( 2\pi \right) ^{2dN}}\,Tr\left( \left( e^{-i\bar{\xi}\eta }e^{-ip\cdot
\bar{x}}\right) \star \left( e^{-\tau L_{0}}\left( e^{i\bar{\xi}\eta
}e^{ip\cdot \bar{x}}\right) \right) \right) .
\end{equation}%
It is a simple exercise to compute it by using the methods above. The result
is given below in Eq.(\ref{oneloop}) where it is in agreement with our next
method of computation in Fourier space. This computation illustrates the
importance of the correct treatment of the anomaly$^{\ref{anom}}$ as will be
emphasized following Eq.(\ref{oneloop}).

\item The tadpole diagram $^{1}$---$\bigcirc $ is obtained from the 3-point
vertex $Tr\left( A_{1}\star A_{2}\star A_{3}\right) $ by identifying legs
2,3. This leads to the expression
\begin{equation*}
\int d^d\bar{x}\int\! \frac{d^dp}{\left( 2\pi \right) ^{d}}\frac{(d\eta) }{%
\left( 2\pi \right) ^{2dN}}\,Tr\left( A_{1}\star \left( e^{-i\bar{\xi}\eta
}e^{-ip\cdot \bar{x}}\right) \star \left( e^{-\tau L_{0}}\left( e^{i\bar{\xi}%
\eta }e^{ip\cdot \bar{x}}\right) \right) \right)
\end{equation*}%
which is again straightforward to evaluate.

\item The one loop correction to the propagator attached to external states $%
_{1}$\_\underline{$\cap $}\_$_{4}$ is obtained by identifying legs 2,3 in
the 4-point function above. This leads to
\begin{align*}
& \int d^d\bar{x}\int\! \frac{d^dp}{\left( 2\pi \right) ^{d}}\frac{(d\eta) }{%
\left( 2\pi \right) ^{2dN}}\,Tr\left( e^{-\tau _{1}L_{0}}\left( A_{1}\star
\left( e^{-i\bar{\xi}\eta }e^{-ip\cdot \bar{x}}\right) \right) \star \left(
e^{-\tau _{2}L_{0}}\left( e^{i\bar{\xi}\eta }e^{ip\cdot \bar{x}}\right)
\right) \star A_{4}\right) \\
& =\left( 2\pi \right) ^{d}\delta^d \left( k_{1}+k_{4}\right) \int\! \frac{%
d^dp}{\left( 2\pi \right)^{d}}\frac{(d\eta) }{\left( 2\pi \right)^{2dN}}%
\,Tr\left( \left( e^{-\tau _{1}L_{0}\left( k_{1}-p\right) }\left( A_{1}\star
e^{-i\bar{\xi}\eta }\right) \right) \star \left( e^{-\tau _{2}L_{0}\left(
p\right) }e^{i\bar{\xi}\eta }\right) \star A_{4}\right)
\end{align*}%
where in the last line the momentum dependent part of $A_{1},A_{4}$ has
already been peeled off, and the $\bar{x}$ integral performed. Then $\beta
_{0}$ in $L_{0}\left( \beta _{0}\right) $ has been replaced by $k_{1}-p$ and
$p$ as appropriate for the propagator with the corresponding momentum.
\end{itemize}

These examples are sufficient to illustrate our approach to such
computations.

\section{Feynman rules in the Fourier basis}

The definition of the action in section 2 is enough to define the Feynman
rules for the open string diagram. Note that in the absence of the last term
in $L_{0}$ of Eq.(\ref{Lo})\ the kinetic term becomes basically the same as
the conventional Lagrangian of the $\phi ^{3}$ theory on the non-commutative
plane \cite{Noncomm}.

\paragraph{Vertex}

The remark above implies that if we take the plane waves $e^{i\bar{\eta} \xi }$ as
the basis to expand the noncommutative field, then the $n$-string
interaction vertex in this basis is,
\begin{equation}
Tr\left( e^{i\bar{\eta _{1}}\xi }\star \cdots \star
e^{i\bar{\eta}_{n}\xi }\right)
=\left(\det{\frac{\sigma}{2\pi}}\right)^{-d/2}\exp \left(
-\frac{1}{2}\sum_{i<j}\bar{\eta}_{i}\sigma \eta
_{j}\right)\delta^{2dN} (\eta _{1}+\cdots +\eta _{n})\,\,,
\label{n-vertex}
\end{equation}%
which is identical to the interaction vertex for non-commutative field
theory.

\paragraph{Propagator}

The simplification of the vertex is compensated by the complication of the
propagator. Now $L_{0}$ is not diagonal. It is still easily computable,
however, by using Eqs.(\ref{evolution}-\ref{tN}) for $M=0,$ ${\cal N}=1$ , $\lambda
=-i\eta ^{\prime },$ and inserting them in Eq.(\ref{prop12})
\begin{eqnarray}
&&\Delta (\eta ,\eta ^{\prime },\tau ,p)\equiv \int \frac{(d\xi )}{(2\pi
)^{2dN}}e^{-i\bar{\eta}\xi }e^{-\tau L_{0}}e^{i\bar{\eta}^{\prime }\xi }
\label{propagator} \\
&&\quad =g(\tau ,p)\exp \left( -\bar{\eta}F(\tau )\eta -\bar{\eta}^{\prime
}F(\tau )\eta ^{\prime }+2\bar{\eta}G(\tau )\eta ^{\prime }+(\bar{\eta}+\bar{%
\eta}^{\prime })H(\tau ,p)\right) ,
\end{eqnarray}%
where%
\begin{eqnarray}
&&g(\tau ,p)=\left( \frac{\theta }{2\pi }\right) ^{dN}(1+\bar{w}w)^{\frac{d}{%
4}}\left( \prod_{e>0}(1-e^{-2\tau \kappa _{e}})\prod_{o>0}(1-e^{-2\tau
\kappa _{o}})\right) ^{-\frac{d}{2}}e^{-\left( {\frac{\tau }{2}}+\bar{w}{%
\frac{\tanh (\frac{\tau \kappa _{e}}{2})}{\kappa _{e}}}w\right)
l_{s}^{2}p^{2}},\quad  \\
&&F(\tau )={1\over 4}M_0^{-1}(\tanh(\tau\tilde{\kappa}))^{-1}
=\left(
\begin{array}{cc}
{\frac{l_{s}^{2}}{2\kappa _{e}}}(\tanh (\tau \kappa _{e}))^{-1} & 0 \\
0 & \frac{\theta ^{2}}{8l_{s}^{2}}\bar{R}\kappa _{o}(\tanh (\tau \kappa
_{o}))^{-1}R%
\end{array}%
\right) , \\
&&G(\tau )={1\over 4}M_0^{-1}(\sinh(\tau\tilde{\kappa}))^{-1}
=\left(
\begin{array}{cc}
{\frac{l_{s}^{2}}{2\kappa _{e}}}(\sinh (\tau \kappa _{e}))^{-1} & 0 \\
0 & \frac{\theta ^{2}}{8l_{s}^{2}}\bar{R}\kappa _{o}(\sinh (\tau \kappa
_{o}))^{-1}R%
\end{array}%
\right) , \\
&&H(\tau ,p)=\frac{\tanh (\tau \kappa _{e}/2)}{\kappa _{e}}wl_{s}^{2}p.
\end{eqnarray}%
A critical difference from the conventional propagator in momentum
representation is that the propagator depends on the momentum variables at
both ends of the propagator in a nontrivial fashion (because momentum is not
conserved due to the potential term in $L_{0}$). Therefore in the Feynman
diagram computation, the momentum integration $d\eta $ is performed at both
ends of each propagator.

\paragraph{External State}

We note also that the external state in Eq.(\ref{external}) is not diagonal
in the momentum basis. We need its Fourier transform
\begin{equation}
\tilde{A}_{\mathcal{N},M,\lambda }=\tilde{\mathcal{N}}e^{-\frac{1}{4}\bar{%
\eta}M^{-1}\eta +\frac{i}{2}\bar{\lambda}M^{-1}\eta }e^{ip\cdot \bar{x}},\;\;%
\tilde{\mathcal{N}}=\mathcal{N}(4\pi )^{-dN}(\det M)^{-d/2}e^{\frac{1}{4}%
\bar{\lambda}M^{-1}\lambda }.
\end{equation}%
For comparison to the oscillator approach, such monoids corresponds to
shifted squeezed states $\exp \left( -\frac{1}{2}a^{\dagger }\mathcal{S}%
a^{\dagger }-\mu a^{\dagger }\right) |p\rangle $, with momentum $p^{\mu }$.
For the general case the relation between oscillator and MSFT parameters is
given in Eqs.(3.4-3.6) of \cite{BM2}. For perturbative states with $\mathcal{%
S}$=0 these reduce to coherent states $\exp (\sum_{n}\mu _{n}\alpha
_{-n})|p\rangle ,$ with a corresponding Moyal field that contains the $M_{0}$
of Eq.(\ref{Mo}) and ${\cal N},\lambda $ given by
\begin{equation}
\mathcal{N}=\left( \det (4\kappa _{o})/\det (\kappa _{e})\right) ^{d/4}~e^{%
\frac{1}{2}(\bar{\mu}_{e}\mu _{e}-\bar{\mu}_{o}\mu _{o})},\;\;\lambda
=i\left(
\begin{array}{c}
-\frac{\sqrt{2}}{l_{s}}\sqrt{\kappa _{e}}\mu _{e}-w_{e}p \\
\frac{2\sqrt{2}il_{s}}{\theta }T\kappa _{o}^{-1/2}\mu _{o}%
\end{array}%
\right) .  \label{lambda-mu}
\end{equation}

To summarize all, the Feynman diagram computation of MSFT reduces to the
following simple prescriptions. As in the conventional field theory, we
decompose the string diagram into the vertex, the propagator and the
external states. We need to perform the momentum integrations attached to
each junction of the components. All integrals are gaussian. Therefore the
computation of any string amplitude reduces to the computation of the
determinant and the inverse of the large matrix which
describes the connections among three components
(the external states, propagator, vertex).
Since the matrices which appear in our computation are explicitly given and
finite dimensional, we obtain a finite and well-defined quantity for
any string diagram.


In order to illustrate our Feynman rule, we present some examples
of the string amplitudes restricted to the matter sector
contribution\footnote{ Although these amplitudes are similar to
some of those in Ref. \cite{Taylor}, our formulas are more general
since they all contain the $ \kappa _{e},\kappa _{o}$ which are
arbitrary frequencies at finite $N$. Furthermore, we can apply
them to nonperturbative external states as they stand, with no
more effort. To obtain the ordinary perturbative string amplitude,
we set $\kappa _{e}=e,\kappa _{o}=o$ and take $N\rightarrow \infty
$ limit (which corresponds to $\bar{w}w\rightarrow \infty $) at
the last stage of computations. }.

\begin{itemize}
\item 1-loop vacuum amplitude

One of the simplest graphs is the 1-loop vacuum amplitude. It can be
computed directly from Eqs.(\ref{evolution}-\ref{tN}), which amounts to
integrating $\Delta (\eta ,\eta ^{\prime },\tau , p)$ (\ref{propagator}):
\begin{eqnarray}
\int d^{d}p\, Tr(e^{-\tau L_{0}\left( p\right) }) &=&\int d^{d}p\int (d\eta)
\Delta (\eta ,\eta ,\tau ,p)  \label{oneloop} \\
&=&(2\pi )^{\frac{d}{2}}l_s^{-d}\tau ^{-{\frac{d}{2}}}\prod_{e>0}(1-e^{-\tau
\kappa_{e}})^{-d}\prod_{o>0}(1-e^{-\tau \kappa _{o}})^{-d},
\end{eqnarray}%
We see that the correct spectrum $\left( \kappa _{e},\kappa _{o}\right) $ is
read off from the 1-loop graph at any $\kappa _{n},N.$ By taking $\kappa
_{e}=e,\kappa _{o}=o$ and $N=\infty $, we reproduce the standard
perturbative string spectrum! Although this graph does not include any
interaction, the coincidence of the spectrum implies the correctness of our
propagator. It is essential to \textit{keep} the term $(1+\bar{w}%
w)^{-1}\left( \sum_{e>0}w_{e}p_{e}\right) ^{2}$ in $L_{0}$ which converts $%
\kappa _{e}$ into $\kappa _{o}$.\footnote{%
This contribution comes from the off-diagonal part of $T\bar{T}$ in Eq.(\ref%
{Lo}).} In fact, if one takes the $\bar{w}w=\infty $ limit \textit{first},
this term drops out and one ends up with the wrong spectrum $\left( \kappa
_{e},\kappa _{e}\right) $ instead of $\left( \kappa _{e},\kappa _{o}\right)
, $ as happened in Ref. \cite{DLMZ}$^{\ref{wrong}}$.

\item 4-tachyon amplitude at the tree level:$_{1}^{2}\rangle _{^{-}5^{---}6^{-}}\langle _{4}^{3} $

As a simple example including interaction, we consider the
perturbative 4-tachyon amplitude $_{1}^{2}\rangle _{^{-}5^{---}6^{-}}\langle _{4}^{3}$ that we discussed
in Eq.(\ref{four}). Following our Feynman rules in the Fourier basis,
we assign the momentum variable $\eta_i$
($i=1,2,\cdots , 6$) to each junctions of the components.
The amplitude is represented by
\begin{eqnarray}
&&\int (d\eta _{1})\cdots (d\eta _{6})e^{-{\frac{1}{2}}(\bar{\eta}_{1}\sigma
\eta _{2}+\bar{\eta}_{2}\sigma \eta _{5}+\bar{\eta}_{1}\sigma \eta _{5})-{%
\frac{1}{2}}(\bar{\eta}_{3}\sigma \eta _{4}+\bar{\eta}_{4}\sigma \eta _{6}+%
\bar{\eta}_{3}\sigma \eta _{6})}  \notag \\
&&\times\delta^{2dN}(\eta _{1}+\eta _{2}+\eta _{5})\delta^{2dN}(\eta
_{3}+\eta _{4}+\eta _{6})\Delta (\eta _{5},-\eta _{6},\tau ,p)\tilde{A}%
_{p_{1}}(\eta _{1})\tilde{A}_{p_{2}}(\eta _{2})\tilde{A}_{p_{3}}(\eta _{3})%
\tilde{A}_{p_{4}}(\eta _{4}),  \notag \\
&&\qquad \tilde{A}_{p_{i}}(\eta )=\mathcal{N}_{0}(4\pi)^{-dN}(\det M_{0})^{-{\frac{d}{2}}}e^{ip_{i}\bar{x}}e^{-{\frac{1}{4}}(\bar{%
\eta}-p_{i}\bar{w})M_{0}^{-1}(\eta -p_{i}w)},
\end{eqnarray}%
where $\tau $ is the length of the propagator and $p=p_{1}+p_{2}-p_{3}-p_{4}$
is the zero mode transfer momentum.
We first perform the momentum integrations over $\eta_5, \eta_6$
to cancel the delta functions which represent the momentum conservations. The remaining integrations are gaussian and
the above expression reduces to
\begin{eqnarray}
&&\mathcal{N}_{0}^{4}{\det {M_{0}}}^{-2d}\left( {%
\frac{2\pi }{4^{4}\theta }}\right) ^{dN}g(\tau ,p)\cdot (\det
\mathcal{A})^{-{\frac{d}{2}}}\cdot e^{{\frac{1}{4}}\bar{\mathcal{B}}\mathcal{%
A}^{-1}\mathcal{B}}e^{-{\frac{1}{4}}(p_{1}^{2}+p_{2}^{2}+p_{3}^{2}+p_{4}^{2})%
\bar{w}M_{0}^{-1}w},
\end{eqnarray}
with
\begin{eqnarray}
&&\mathcal{A}=\left(
\begin{array}[tb]{cccc}
F(\tau )+{\frac{1}{4}}M_{0}^{-1} & F(\tau )+{\frac{1}{4}}\sigma  &
G(\tau )
& G(\tau ) \\
F(\tau )-{\frac{1}{4}}\sigma  & F(\tau )+{\frac{1}{4}}M_{0}^{-1} &
G(\tau )
& G(\tau ) \\
G(\tau ) & G(\tau ) & F(\tau )+{\frac{1}{4}}M_{0}^{-1} & F(\tau )+{\frac{1}{4%
}}\sigma  \\
G(\tau ) & G(\tau ) & F(\tau )-{\frac{1}{4}}\sigma  & F(\tau )+{\frac{1}{4}}%
M_{0}^{-1}
\end{array}%
\right) , \label{VeneA}\\
&&\bar{\mathcal{B}}=\left( -\bar{H}+{\frac{1}{2}}p_{1}\bar{w}M_{0}^{-1},-%
\bar{H}+{\frac{1}{2}}p_{2}\bar{w}M_{0}^{-1},\bar{H}+{\frac{1}{2}}p_{3}\bar{w}%
M_{0}^{-1},\bar{H}+{\frac{1}{2}}p_{4}\bar{w}M_{0}^{-1}\right) .
\label{VeneB}
\end{eqnarray}%
In the matrix $\mathcal{A}$, $M_{0}$ comes from the external tachyons, $F,G$
from the propagator and $\sigma $ from the Moyal $\star $ product. On the
other hand $\bar{\mathcal{B}}$ originates from zero-mode
momentum-dependent terms. The appearances of $F$ at the off-diagonal
parts in ${\cal A}$ come from the mixing induced from
the momentum integrations over $\eta_{5,6}$.

This formula looks more complicated than the expression given
in Eq.(\ref{4-amp_rev}) since it involves the matrices of larger
size.  The reduction to (\ref{4-amp_rev}) can be proved
by the reorganization of the gaussian integrations which is
outlined in the next section.

\item 2-loop vacuum amplitude:\
$\overset{1}{\underset{3}{_{2}}}\langle
\frac{^{---~b~--}}{{_{---~a~--}}} \rangle
\overset{5}{\underset{4}{_{6}}}$

As a next example,
we consider one of the 2-loop vacuum graphs:
Two 3-string vertices are connected by three propagators which have
momenta $p_{a},p_{b},p_{c}$ and lengths $\tau _{a},\tau _{b},\tau _{c}$.
We assign the momentum variables $\eta_i$ ($i=1,\cdots,6$)
as depicted in the figure.
\begin{eqnarray}
&&\int (d\eta _{1})\cdots (d\eta _{6})e^{-{\frac{1}{2}}(\bar{\eta}_{1}\sigma
\eta _{2}+\bar{\eta}_{1}\sigma \eta _{3}+\bar{\eta}_{2}\sigma \eta _{3})-{%
\frac{1}{2}}(\bar{\eta}_{4}\sigma \eta _{6}+\bar{\eta}_{4}\sigma \eta _{5}+%
\bar{\eta}_{6}\sigma \eta _{5})}  \notag \\
&&\qquad\times\delta^{2dN} (\eta _{1}+\eta _{2}+\eta _{3})\delta^{2dN}(\eta
_{4}+\eta _{5}+\eta _{6})  \notag  \label{2loop} \\
&&\qquad \times \Delta (\eta _{3},-\eta _{4},\tau _{a},p_{a})\Delta (\eta
_{1},-\eta _{5},\tau _{b},p_{b})\Delta (\eta _{2},-\eta _{6},\tau _{c},p_{c})
\notag \\
&=&g(\tau _{a},p_{a})g(\tau _{b},p_{b})g(\tau_{c},p_{c}) \left( \det \left(
\pi^{-1}\mathcal{M}\right) \right) ^{-{\frac{d}{2}}}\int {%
\frac{(d\chi _{1})}{(2\pi)^{2dN} }}{\frac{(d\chi _{2})}{(2\pi)^{2dN} }}e^{{%
\frac{1}{4}}\bar{\mathcal{K}}\mathcal{M}^{-1}\mathcal{K}}
\label{two-loop}
\end{eqnarray}%
where, after denoting $H(\tau_{a},p_{a})$ as $H_{a}$, the quantities $\mathcal{M},\mathcal{K}$ are given by
\begin{eqnarray}
\mathcal{M} &=&\left(
\begin{array}[tb]{cccccc}
F(\tau _{b}) & {\frac{1}{4}}\sigma & {\frac{1}{4}}\sigma & 0 & G(\tau _{b})
& 0 \\
-{\frac{1}{4}}\sigma & F(\tau _{c}) & {\frac{1}{4}}\sigma & 0 & 0 & G(\tau
_{c}) \\
-{\frac{1}{4}}\sigma & -{\frac{1}{4}}\sigma & F(\tau _{a}) & G(\tau _{a}) & 0
& 0 \\
0 & 0 & G(\tau _{a}) & F(\tau _{a}) & {\frac{1}{4}}\sigma & {\frac{1}{4}}%
\sigma \\
G(\tau _{b}) & 0 & 0 & -{\frac{1}{4}}\sigma & F(\tau _{b}) & -{\frac{1}{4}}%
\sigma \\
0 & G(\tau _{c}) & 0 & -{\frac{1}{4}}\sigma & {\frac{1}{4}}\sigma & F(\tau
_{c})
\end{array}%
\right) , \\
\bar{\mathcal{K}} &=&\left( i\bar{\chi _{1}}+\bar{H}_{b},i\bar{%
\chi _{1}}+\bar{H}_{c},i\bar{\chi _{1}}+\bar{H}_{a},i\bar{\chi _{2}}-\bar{H}%
_{a},i\bar{\chi _{2}}-\bar{H}_{b},i\bar{\chi
_{2}}-\bar{H}_{c}\right) . \label{K}
\end{eqnarray}%
In passing from the first to the second line in Eq.(\ref{two-loop}),
we rewrote the delta functions at the vertices by,
\begin{equation}
\label{trick}
\delta^{2dN}(\eta_1+\eta_2+\eta_3)=\int\frac{(d\chi_1)}{(2\pi)^{2dN}}
e^{i\bar\chi_1(\eta_1+\eta_2+\eta_3)}\,\,,
\end{equation}
and performed the $\eta_i$ integrations.
\end{itemize}

We are now ready to write down the explicit form of the
scattering amplitude for any Feynman diagram.
We use the trick Eq.(\ref{trick}) to convert the delta
functions at the vertices to the gaussian integration
over $\chi$s.  The general formula after $\eta$ integrations
is given by
\begin{eqnarray}
\prod_{a\in\left\{{\rm propagators}\right\}}
g(\tau_a,p_a)\cdot \prod_{u\in\left\{{\rm external\ states}\right\}}{\tilde {\cal N}_u}e^{ip_u\bar{x}}\cdot \left(\det\left(\pi^{-1}{\cal M}\right)\right)^{-{d\over 2}}\cdot
\int \prod_{i\in\left\{{\rm vertices}\right\}}{(d\chi_i)\over (2\pi)^{2dN}}e^{{1\over 4}\bar{\cal K}{\cal M}^{-1}{\cal K}}
\label{universal}
\end{eqnarray}
where the matrix ${\cal M}$ is
\begin{eqnarray}
 {\cal M}_{II}&=&\left\{
\begin{array}[tb]{cl}
 F(\tau_a) & I\in\textrm{vertex}(i)\cap\textrm{propagator}(a)\\
 {1\over 4}M_u^{-1} & I\in\textrm{vertex}(i)\cap
\textrm{external state}(u)\\
F(\tau_a)+{1\over 4}M_u^{-1} & I\in\textrm{propagator}(a)
\cap\textrm{external state}(u)
\end{array}
\right.\quad ,\label{mii}\\
 {\cal M}_{IJ}&=&\left\{
\begin{array}[tb]{cl}
\pm {1\over 4}\sigma & I,J\in \textrm{vertex}(i) \\
\pm G(\tau_a) & I,J\in \textrm{propagator}(a)\\
0 & \textrm{otherwise}
\end{array}
\right.\qquad (I\ne J)\, ,
\label{mij}
\end{eqnarray}
and vector ${\cal K}$ is
\begin{eqnarray}
 {\cal K}_I=\left\{
\begin{array}[tb]{cl}
i\chi_i\pm H(\tau_a,p_a) &I\in\textrm{vertex}(i)
\cap\textrm{propagator}(a)\\
i\chi_i+{i\over 2}M_u^{-1}\lambda_u &I\in\textrm{vertex}(i)
\cap\textrm{external state}(u)\\
0 & \text{otherwise}
\end{array}
\right.\,.\label{ki}
\end{eqnarray}
The indices $I,J,\cdots$ in the matrix ${\cal M}$ and ${\cal K}$
represent the junctions between the arbitrary combinations of the
basic components (propagators, external states and vertices). In
Eqs.(\ref{mii},\ref{mij}), vertex$(i)$ (propagator$(a)$, external
state $(u)$) represents the set of the boundaries of the
$i^{\textrm{th}}$ vertex ($a^{\textrm{th}}$ propagator,
$u^{\textrm{th}}$ external state). We specify the junction $I$ by
taking the common set among them as indicated by a Feynman graph.
In the example of the two-loop amplitude, all the indices
$I=1,\cdots,6$ describe the junction between a vertex and a
propagator. Applying the rules we obtain
Eqs.(\ref{two-loop}-\ref{K}). Similarly, in the example of the
4-tachyon scattering, the indices $I=1,2,3,4$ describe the
vertex-external state junctions while the indices $I=5,6$ describe
the vertex-propagator junctions. In this example, after applying
these rules we first construct a $6\times6$ matrix $\mathcal{M}$
and a corresponding $\mathcal{K}$. After integrating over
$\chi_{1,2}$ we obtain the $4\times 4$ matrix $\mathcal{A}$ and
the corresponding $\mathcal{B}$ given in Eqs.(\ref{VeneA},
\ref{VeneB}).
An example of propagator-external state junction appears in the
Feynman graph considered in Eq.(\ref{prop12}) in the
previous setion\footnote{%
Strictly speaking there are other possibilities for
the junctions, for instance, propagator-propagator, vertex-vertex.
However, since they can be obtained by taking the appropriate
limit (for example infinitely short propagator), we do not write
them explicitly.}.
%
%

The plus or minus signs in Eqs.(\ref{mij},\ref{ki}) are determined
by the relative positions of the labels $I$ in the diagram. For
example, the sign of $\mathcal{M}_{IJ}\rightarrow \pm {1\over
4}\sigma$ at a vertex is $+(-)$ if $I$ is positioned after 
(before) $J$ when going clockwise around the vertex.
To figure out the signs of $G,H$ systematically, the diagram may 
be decorated with arrows for all momenta directed into each 
vertex. Recall that each propagator has different momenta at each 
end, therefore a propagator with ends $(I,J)$ will have momenta 
$(+\eta_I,-\eta_J)$. The sign in front of $G$ is given by the 
product of the signs of the momenta at the two ends of the 
propagator, times $(-1)$. So if the arrows are drawn as suggested, 
the sign is $+G$. Finally the sign in front of $H$ is determined by 
the sign of $\eta_I$ at each end of the propagator. If the 
direction of the arrows is changed according to some other 
convention the signs on $G,H$ will flip accordingly.

\section{Reorganization of gaussian integration}


Our computation of Feynman diagrams in Fourier basis reduces to the
computation of the determinant and the inverse of the large matrices which
connect all the external states. It is somehow obscure how such computation
is related to the computation in $\xi$ basis presented in section 3. To see
the correspondence more explicitly, it is illuminating to carry out some of
the momentum integrals.

For that purpose, we dissect all the propagators which connect two vertices.
In the following, we carry out the momentum integrations associated with
each vertex attached to the dissected propagators. More explicitly, we
consider the following integration,

\begin{eqnarray}
V_{n}(\tau ,\epsilon ) &\equiv &\int (d\eta _{1})\cdots (d\eta
_{n})\,\delta^{2dN}(\eta _{1}+\cdots +\eta _{n})\,e^{-\frac{1}{2}\sum_{i<j}%
\bar{\eta}_{i}\sigma \eta _{j}}  \notag \\
&&\times \prod_{i=1}^{n}\left( g(\tau _{i},p_{i})\,e^{-\bar{\eta}_{i}F(\tau
_{i})\eta _{i}+\bar{\epsilon}_{i}G(\tau _{i})\eta _{i}+(\bar{\eta}_{i}+\bar{%
\epsilon}_{i})H(\tau _{i},p_i)-\bar{\epsilon _{i}}F(\tau _{i})\epsilon
_{i}}\right) \,\,.
\end{eqnarray}%
The first line comes from the definition of the vertex in the momentum basis
and the second line comes from the propagator. We note that the second line
may be written as, $\prod_{i}\tilde{A}_{\tilde{\mathcal{N}}_i,\tilde{M_{i}},%
\tilde{\lambda}_{i}}$ where
\begin{equation}
\tilde{A}_{\tilde{\mathcal{N}}_i,\tilde{M_{i}},\tilde{\lambda}_{i}}\equiv
\tilde{\mathcal{N}}_i e^{-\bar{\eta} \tilde{M}_i\eta -\bar{\eta}\tilde{\lambda}_i },\quad
\tilde{\mathcal{N}}_{i}=g(\tau _{i},p_{i})e^{\bar{\epsilon}_{i}H(\tau
_{i},p_i)-\bar{\epsilon}_{i}F(\tau _{i})\epsilon _{i}},\,\tilde{M}%
_{i}=F(\tau _{i}),\,\tilde{\lambda}_{i}=-G(\tau _{i})\epsilon _{i}-H(\tau
_{i},p_i)\,.
\end{equation}%
Actually this is nothing but the trace of the Moyal product of $n$ gaussian
functions,
\begin{eqnarray}
V_{n}(\tau ,\epsilon ) &=&Tr\left( A_{\mathcal{N}_{1},M_{1},\lambda
_{1}}(\tau _{1},\epsilon _{1})\star \cdots \star A_{\mathcal{N}%
_{n},M_{n},\lambda _{n}}(\tau _{n},\epsilon _{n})\right) , \\
M_{i} &=&\left(4F(\tau _{i})\right)^{-1}\,\,,\quad \lambda _{i}=-\frac{i}{2}%
F(\tau_{i})^{-1}(G(\tau _{i})\epsilon _{i}+H(\tau _{i},p_i)) , \\
\mathcal{N}_{i} &=&g(\tau _{i},p_{i})\pi^{dN}(\det F(\tau _{i}))^{\frac{d}{2}%
}\,e^{\bar{\epsilon}_{i}H(\tau _{i},p_i)-\bar{\epsilon}_{i}F(\tau
_{i})\epsilon _{i}+\frac{1}{4}(\bar{H}(\tau _{i},p_i)+\bar{\epsilon}%
_{i}G(\tau _{i}))F(\tau _{i})^{-1}(H(\tau _{i},p_i)+G(\tau _{i})\epsilon )}.
\notag
\end{eqnarray}%
While it looks complicated, the explicit evaluation of such expressions are
given in \cite{BM2}.

We note that once this expression is evaluated, one may write down any
string amplitude schematically as,
\begin{equation}
\mathcal{A}\sim \int (d\epsilon) \prod_{\mathrm{vertices}}V_{n^{(i)}}(\tau
^{(i)},\epsilon ^{(i)})\prod_{\mathrm{external\,\,legs}}\!\tilde{A}_{\tilde{\mathcal{N}_i},\tilde{M}_i,\tilde{\lambda}_i}\prod_{\mathrm{connection}%
}\!\delta^{2dN}(\epsilon _{i}+\epsilon _{j})
\end{equation}%
where the final factor describes the momentum integrations for each
dissection point of the propagators. The second factor comes from the
external states. As an example, the two-loop vacuum amplitude is now neatly
written as,
\begin{equation}
\int (d\epsilon _{1})(d\epsilon _{2})(d\epsilon _{3})V_{3}(\tau _{1},\tau
_{2},\tau _{3};\epsilon _{1},\epsilon _{2},\epsilon _{3})V_{3}(\tau
_{1}^{\prime },\tau _{2}^{\prime },\tau _{3}^{\prime };-\epsilon
_{1},-\epsilon _{2},-\epsilon _{3}).
\end{equation}%
Other amplitudes can be written as easily as this one.

\section{Moyal formulation of $bc$-ghost sector}

In Ref.\cite{BM2}, the Moyal formulation of \textit{bosonized} ghost was
discussed and was shown to be almost the same as a matter boson. In certain
explicit computations in the ghost sector, it is often convenient to use the
fermionic $b,c$-ghosts. Because $b(\sigma ),c(\sigma )$ have $\cos n\sigma $
as well as $\sin n\sigma $ modes, we need to develop a regularized version
of half string formulation for sine modes.\footnote{%
The cosine modes were developed in Ref.\cite{BM1}\cite{BM2}. This was enough
to discuss matter and bosonized ghost sector. The half-string formulation of
$b,c$ ghost was developed in \cite{Bord}.}. In the ordinary split string
formulation, we find the infinite matrix
\begin{equation}
{\tilde{R}}_{oe}={\frac{4}{\pi }}\int_{0}^{\frac{\pi }{2}}d\sigma \sin
o\sigma \sin e\sigma ={\frac{4e(i)^{e-o+1}}{\pi (e^{2}-o^{2})}}.
\label{infiniteR}
\end{equation}%
The inverse matrix is its transpose: ${\tilde{R}}\bar{\tilde{R}}=1_o,\bar{\tilde{R}}{\tilde{R}}=1_e$. However, ${\tilde{R}}$ has a zero mode $\tilde{w}_{o}=%
\sqrt{2}(i)^{o-1},$ namely $\sum_{o=1}^{\infty }\tilde{w}_{o}{\tilde{R}}%
_{oe}=0$. As emphasized in footnote (\ref{anom}), this situation causes an
associativity anomaly of infinite matrices, which leads to ambiguities in
computation as was discussed in Ref.\cite{BM1}. To perform well-defined
computations, we construct a finite $N\times N$ matrix ${\tilde{R}}$ using
arbitrary spectrum $\kappa _{e},\kappa _{o}$ as we did for $T,R,w,v$ in \cite%
{BM2}
\begin{equation}
{\tilde{R}}_{oe}:={\frac{w_{e}v_{o}\kappa _{e}\kappa _{o}}{\kappa
_{e}^{2}-\kappa _{o}^{2}}}.
\end{equation}%
The original $\tilde{R}$ (\ref{infiniteR}) is recovered by putting $\kappa
_{e}=e,\kappa _{o}=o$ and taking $N\rightarrow \infty $.

We now follow a procedure parallel to that in \cite{B}. Using the finite
version of $T,R,\tilde{R},v,w$, we define half string modes for $b(\sigma
),c(\sigma )$, and perform the Fourier transform with respect to the even
modes of the full string:\footnote{%
In the matter sector the Moyal product could be formulated in either the
even or odd sectors \cite{B}\cite{moyalMoscow}, in the $bc$ sector it is
more natural to formulate it using odd modes.}
\begin{eqnarray}
&&A(\xi _{0},x_{o},p_{o},y_{o},q_{o}):=\int
dc_{0}\prod_{e>0}(dx_{e}dy_{e})\,e^{-\xi _{0}c_{0}+(\xi _{0}+{\frac{2}{g}}%
q_{o}v)\bar{w}y_{e}+{\frac{2}{g}}p_{o}{\tilde{R}}x_{e}+{\frac{2}{g}}q_{o}%
\bar{T}y_{e}}\langle c_{0},x_{n},y_{n}|\Psi \rangle , \\
&&\langle c_{0},x_{n},y_{n}|=\langle \Omega |{\hat{c}}_{-1}{\hat{c}}_{0}\exp
\left( c_{0}\hat{b}_{0}+\sum_{n>0}\left( -\hat{c}_{n}\hat{b}_{n}-i\sqrt{2}%
\hat{c}_{n}x_{n}+\sqrt{2}y_{n}\hat{b}_{n}+iy_{n}x_{n}\right) \right) .
\end{eqnarray}%
The \textit{Grassmann odd} version of the Moyal product among them\footnote{%
The Moyal formulation of the ghost system was discussed in
Ref.\cite{Erler} in the context of the continuous basis. The
author also defined the discrete Moyal star product using {\it
even} modes starting from the continuous product. While our switch
to odd modes appears trivial, there are some important differences
between our treatment and that of \cite{Erler}. A critical point
is the treatment of the midpoint mode $b(\pi/2)$ which is
nontrivial in our case but vanishes in \cite{Erler}. This has an
important consequence for producing the correct Neumann
coefficients. Actually we also developed the {\em even} mode
formulation as \cite{Erler} but with the proper treatment of the
midpoint. It is, however, more complicated than the odd mode
formulation presented here. Another important difference is on the
regularization of the infinite matrices where our setup holds for
arbitrary $\kappa_e,\kappa_o,N$.
The details will be given in \cite{PREP}. Also for more comments
on midpoint issues related to the continuous basis, see
\cite{moyalMoscow}.}:
\begin{equation}
A\star B=A\exp \left( {\frac{g}{2}}\sum_{o>0}\left( {\frac{\overleftarrow{%
\partial }}{\partial x_{o}}}{\frac{\overrightarrow{\partial }}{\partial p_{o}%
}}+{\frac{\overleftarrow{\partial }}{\partial y_{o}}}{\frac{\overrightarrow{%
\partial }}{\partial q_{o}}}+{\frac{\overleftarrow{\partial }}{\partial p_{o}%
}}{\frac{\overrightarrow{\partial }}{\partial x_{o}}}+{\frac{\overleftarrow{%
\partial }}{\partial q_{o}}}{\frac{\overrightarrow{\partial }}{\partial y_{o}%
}}\right) \right) B
\end{equation}%
corresponds to (anti-)overlapping condition of Witten's star product: $%
b^{(r)}(\sigma )-b^{(r-1)}(\pi -\sigma )=0,\ c^{(r)}+c^{(r-1)}(\pi -\sigma
)=0$. Here $x_{n},y_{n},p_{o},q_{o},c_{0},\xi _{0}$ are Grassmann odd
variables. A ghost zero mode $\xi _{0}$ also enters in reproducing Witten's
star along with the above Moyal $\star $ product. Through the oscillator
formalism \cite{GJ2} we establish the link between Witten's product and our
Moyal product as follows
\begin{equation}
\int d\xi _{0}^{(1)}d\xi _{0}^{(2)}d\xi _{0}^{(3)}~Tr\left[ A^{(1)}(\xi
_{0}^{(1)},\xi )\star A^{(2)}(\xi _{0}^{(2)},\xi )\star A^{(3)}(\xi
_{0}^{(3)},\xi )\right] \sim \langle \Psi ^{(1)}|\langle \Psi ^{(2)}|\langle
\Psi ^{(3)}|V_{3}\rangle ,
\end{equation}%
where
\begin{equation}
\xi =(x_{o},p_{o},y_{o},q_{o}),\quad TrA(\xi ):=\int
\prod_{o>0}(dx_{o}dp_{o}dy_{o}dq_{o})\,A(\xi ).
\end{equation}%
In fact, by substituting the coherent states and their Fourier transform for
$\Psi ^{(i)},A^{(i)}$, we have verified that the Neumann coefficients in the
above identification coincide with the ones which were defined using
6-string vertex in \textit{matter} sector in Ref.\cite{BM2}. This provides a
successful test of the ghost zero mode part.\footnote{%
The coincidence of the Neumann coefficients for nonzero mode implies that
our Moyal $\star $ product is essentially the same as the \textit{reduced}
product which was introduced in the Siegel gauge \cite{IK,Oku}.} This
coincidence holds for arbitrary $\kappa _{e},\kappa _{o},N$. As usual, we
reproduce the ordinary Neumann coefficients of Witten's string field theory
by putting $\kappa _{e}=e,\kappa _{o}=o$ and taking $N\rightarrow \infty $
in the last stage of computations.

\section{Comment on relation with VSFT}

So far, we have established the utility of MSFT for computing Feynman graphs
with perturbative as well as nonperturbative external states. To make
further progress with nonperturbative effects, it will be important to
understand how MSFT could be used in vacuum string field theory. In this
section, we make some remarks in this direction.

The second order differential operator $L_{0}\left( \beta _{0}\right) ,$
including the matter and ghost sectors, can be rewritten using the Moyal $%
\star $ product as follows
\begin{equation*}
L_{0}A=\mathcal{L}_{0}(\beta _{0})\star A+A\star \mathcal{L}_{0}(-\beta
_{0})+\gamma A,
\end{equation*}%
where%
\begin{eqnarray}
\mathcal{L}_{0}(\beta _{0}) &=&\sum_{e>0}\left( {\frac{l_{s}^{2}}{\theta ^{2}%
}}p_{e}^{2}+{\frac{\kappa _{e}^{2}}{4l_{s}^{2}}}x_{e}^{2}-{\frac{l_{s}}{%
\theta }}w_{e}p_{e}\beta _{0}\right) +{\frac{1}{4}}(1+\bar{w}w)\beta
_{0}^{2}-{\frac{d-2}{4}}\sum_{n>0}\kappa _{n}  \notag \\
&&+i\sum_{o>0}\kappa _{o}\left( \frac{1}{2}x_{o}y_{o}+{\frac{2}{g^{2}}}p_{o}q_{o}\right) , \\
\gamma &=&-{\frac{1}{1+\bar{w}w}}{\frac{2l_{s}^{2}}{\theta ^{2}}}\left(
\sum_{e>0}w_{e}p_{e}\right) ^{2}+{\frac{4i}{g^{2}}}(1+\bar{w}w)\left(
\sum_{o>0}\kappa _{o}v_{o}p_{o}\right) \left( \sum_{o>0}v_{o}q_{o}\right) .
\end{eqnarray}%
The $\mathcal{L}_{0}$ terms are star products with a field without involving
explicit derivatives with respect to $\xi .$ However, the $\gamma $ term is
an ordinary product, not a star product. It involves $\sum_{e>0}p_{e}^{\mu
}w_{e}$ which can be rewritten as%
\begin{equation}
\sum_{e>0}p_{e}^{\mu }w_{e}=\sum_{e,o>0}p_{o}^{\mu }R_{oe}w_{e}=\left( 1+%
\bar{w}w\right) \sum_{o>0}p_{o}^{\mu }v_{o}=\left( 1+\bar{w}w\right) \tilde{p%
}^{\mu },
\end{equation}%
where the mode $\tilde{p}^{\mu }\equiv \sum_{o>0}v_{o}p_{o}^{\mu }$ was
discussed in \cite{BM1} as being closely related to associativity anomalies
in string field theory. Also note that in the large $N$ limit $\tilde{p}%
^{\mu }$ becomes the \textit{unpaired} zero momentum mode in the continuum
Moyal representation of ref.\cite{DLMZ}: $\left( \lim_{k\rightarrow 0}p^{\mu
}\left( k\right) \right) \sim \tilde{p}^{\mu }.$ The ghost part has a
similar structure. Evidently, these bits are closely connected to midpoint
anomalies.

If we could neglect the $\gamma $ term, $L_{0}A$ would be given by
left $\star $ multiplication with $\mathcal{L}_{0}(\beta _{0})$
and right star multiplication with $\mathcal{L}_{0}(-\beta _{0})$.
The left-right splitting of the kinetic term reminds us of the
situation of the purely cubic string field theory \cite{PSFT}
where the BRST operator $Q_B$ was decomposed into the left and
right star multiplication of the string fields $Q_L I$ or $Q_R I$.
In this sense, purely cubic theory
 is essentially the matrix analog of Witten's string
field theory. In our MSFT framework $Q_B,Q_L I$ correspond to
$L_0$ and ${\cal L}_0$ respectively because we are in the Siegel
gauge. However, we have now seen that this structure must be
corrected with our $\gamma$ term.

One of the lessons we learned in this paper is that we can not
neglect the gamma term because it is indispensable to reproduce
the correct spectrum in the computation of 1-loop vacuum
amplitude, and other quantities, in both matter and ghost sectors,
even if the coefficients in front of them appear to vanish naively
in the large $N$ limit.

The origin of the $\gamma$ term is a non-vanishing energy-momentum
tensor at the midpoint.  While the integration measure is zero, it
still gives a finite contribution. The situation is similar in the
gauge covariant BRST formulation, namely the BRST current does not
vanish at the midpoint. Our observation here indicates that one
might need a more careful analysis of the midpoint BRST operator
in the purely cubic theory and/or in VSFT.

We are currently in the process of solving the classical equations of
motion of the original theory\footnote{%
There are some attempts to solve it in different scheme
\cite{solution}.} and hope to establish a careful connection
between VSFT and the original theory in the context of MSFT. We
expect that the more careful treatment of this term would clarify
the midpoint structure of VSFT. The finite $N$ regularization
which is used in this paper will be essential in this viewpoint.

\section{Outlook}

In this paper, we have restricted ourselves to computations in the Siegel
gauge, and demonstrated the utility of MSFT.

Ideally, for further insights, we would like to aim for a more gauge
invariant approach. To achieve this carefully it is necessary to construct
the nilpotent BRST operator. We can do this in the infinite $N$ limit, but
not yet in the finite $N$ case. The reason is that the Virasoro algebra does
not close with a finite number of modes. Therefore one needs to find a
finite dimensional algebra that closes at finite $N,$ and becomes the
Virasoro algebra at infinite $N.$ With such an algebra one can construct a
BRST operator and a gauge invariant Lagrangian at finite $N.$ This would be
the analog of lattice gauge theory for QCD. The cutoff theory we have
described so far would correspond to the gauge fixed lattice gauge theory.
It is possible that in this process the $\left( \kappa _{e},\kappa
_{o}\right) $ that have remained arbitrary so far in our formalism would be
fixed as a function of $\left( e,o\right) $ at finite $N.$

On the other hand, if the VSFT proposal is valid, we can easily construct
the midpoint nilpotent BRST operator from only midpoint degrees of freedom.
In that context a gauge invariant theory is easily constructed without ever
encountering a restriction on the $\left( \kappa _{e},\kappa _{o}\right) $
at finite $N.$

The clarification of such issues will be critical for the future development
of string field theory.

\begin{center}
\noindent{\large \textbf{Acknowledgments}}
\end{center}

I.B. is supported in part by a DOE grant DE-FG03-84ER40168. I.K. is
supported in part by JSPS Research Fellowships for Young Scientists. Y.M. is
supported in part by Grant-in-Aid (\# 13640267) from the Ministry of
Education, Science, Sports and Culture of Japan.


\end{document}